\documentclass[aps,pra,preprint,groupedaddress]{revtex4-1}

\usepackage{graphics}
\usepackage{graphicx}
\usepackage{longtable}
\begin{document}



\title{Spectroscopy of diagnostically-important magnetic-dipole
lines in highly-charged 3d$^n$ ions of tungsten}


\author{Yu.~Ralchenko}
\email[Electronic mail: ]{yuri.ralchenko@nist.gov}
\author{I.N.~Dragani\'{c}}
\altaffiliation{Current address: Oak Ridge National Laboratory, Oak Ridge TN 37831-6372}
\author{D.~Osin}
\author{J.D.~Gillaspy}
\author{J.~Reader}
\affiliation{National Institute of Standards and Technology, Gaithersburg,
Maryland 20899-8422}


\date{\today}

\begin{abstract}

An electron beam ion trap (EBIT) is used to measure extreme ultraviolet
spectra between 10~nm and 25~nm from highly-charged ions of tungsten with an
open $3d$ shell (W XLVIII through W LVI). We found that almost all strong
lines are due to the forbidden magnetic-dipole (M1) transitions within $3d^n$
ground configurations. A total of 37 spectral lines are identified for the
first time using detailed collisional-radiative (CR) modeling of the EBIT
spectra. A new level-merging scheme for compactification of rate equations is
described. The CR simulations for Maxwellian plasmas show that several line
ratios involving these M1 lines can be used to reliably diagnose temperature
and density in hot fusion devices.

\end{abstract}


\maketitle



\section{Introduction}

Spectroscopy of highly-charged ions of high-Z elements is currently the
subject of extensive research. From a theoretical viewpoint, the accurately
measured wavelengths, energy levels and transition probabilities provide
crucial tests for advanced theories of atomic structure in a regime where
relativistic and quantum-electrodynamic effects become very strong. As for
applications, since tungsten is currently considered to be a primary candidate
for the plasma-facing material in the ITER divertor region \cite{Hawryluk09},
the spectra of its ions in a wide range of wavelengths are being studied under
various conditions. It is not surprising, therefore, that a large number of
research papers on the spectra of tungsten ions measured with electron beam
ion traps (EBIT)
\cite{11850EL,9743EL,11159EL,9064EL,8472EL,9000EL,12286EL,12366EL,8219TP,14870EL,
PhysRevA.81.012505,15007EL}, tokamaks \cite{12101EL,14670EL,15280EL},
stellarators \cite{15343EL} and other high-temperature-plasma devices were
published over the last decade. A detailed compilation of the recent results
on spectral lines and spectra of W can be found in
Refs.~\cite{12177EL,14671EL}.

In the ITER plasma, the tungsten ions will be transported from the relatively
cold divertor region to the plasma core with temperatures on the order of
20~keV. Although considerable efforts are to be spent to minimize radiative
power losses due to emission from highly-charged ions of W, very useful
information for plasma diagnostics can be derived from isolated spectral
lines. For instance, the electron temperature $T_e$ can be easily found from
the ratios of strong lines from different ions through the dependence of the
ionization balance on $T_e$, and the ion temperature can be derived from the
line shapes. Determination of the electron density $n_e$ from spectral lines,
however, is not as straightfoward. Most often it involves a comparison of
allowed and forbidden lines, and thus this technique relies upon knowledge of
wavelengths and transition probabilities of the involved spectral lines. At
high densities, when level populations approach the local thermodynamic
equilibrium (LTE), or Boltzmann, limit, forbidden lines with transition
probabilities many orders of magnitude smaller than those for allowed
electric-dipole (E1) transitions are too weak to be observed in the spectrum.
In low density plasmas, however, the populations of the excited levels which
decay only via forbidden transitions can be relatively high and therefore
result in strong intensities. For each forbidden line there exists a
transition range of electron densities where electron collisions are
comparable to the radiative decay rate. It is in this range of densities
that one may hope to use a particular forbidden line for density diagnostics.

Currently, more than 80 forbidden lines in tungsten ions, from W$^{28+}$ to
W$^{57+}$, are known from experimental measurements \cite{ASD}. The
high-multipole lines from W ions were observed, for instance, in x-rays
\cite{9823EL,10172EL, 11848EL, 9000EL, PhysRevA.81.012505}, extreme
ultraviolet (EUV) \cite{12101EL, 8219TP,12366EL}, vacuum ultraviolet
\cite{12340EL} and ultraviolet (UV) \cite{11850EL,9743EL} ranges of spectra in
tokamaks and EBITs. The probabilities of forbidden transitions show very
strong increase with the ion charge while the collisional damping of spectral
lines becomes less effective due to a decrease of cross sections. As a result,
the forbidden lines are more prominent in the spectra of multiply-charged
ions.

Forbidden transitions in highly-charged ions are also a subject of active
theoretical research with emphasis on their use in plasma diagnostics.  The
visible/UV magnetic-dipole (M1) $J=2-3$ line in Ti-like ions was analyzed for
density diagnostics in hot plasmas since the pioneering work of Feldman et al.
\cite{10637EL}. Continuing this work, Feldman et al. \cite{7416TP,11893EL}
performed a systematic study of density-sensitive M1 lines in Ti-like ions and
in various N-shell ions. Recently Jonauskas et al. \cite{8699TP} calculated
wavelengths and transition probabilities of M1 lines in $4d^n$ configurations
of W ions using large-scale configuration-interaction methods. Also, Quinet et
al. \cite{8768TP} performed Hartree-Fock calculations of allowed and forbidden
transitions in W I--III, addressing in particular the diagnostics of fusion
plasmas. An extensive calculation of atomic characteristics of eight
isoelectronic sequences of tungsten ions in a broad range of wavelengths and
transitions was recently performed by Safronova and Safronova using the
relativistic many-body perturbation theory (RMBPT) \cite{8725TP}.  The number
of theoretical works on spectral lines and transition probabilities in ions of
tungsten is too large to cite here, so we refer the reader to the
bibliographic databases at the National Institute of Standards and Technology
(NIST) for an extensive list of publications on tungsten \cite{NISTbib}.

An example of a density-sensitive line ratio in highly-charged tungsten ions
can be provided by the ratio of electric-quadrupole (E2) and magnetic-octupole
(M3) lines in Ni-like W$^{46+}$ \cite{12263EL}. These two close lines  at
about 0.793 nm are due to two $3d^{10}$--$3d^94s$ parity-conserving
transitions. They were experimentally resolved only recently
\cite{PhysRevA.81.012505}, although the unresolved spectral feature was known
for several years from tokamak \cite{9823EL} and EBIT \cite{9000EL}
measurements. A large difference in transition probabilities for these lines,
on the order of $10^6$, results in a different response to collisional
destruction of level populations. As was shown in Ref.~\cite{12263EL}, the
E2/M3 line ratio in W$^{46+}$ can be used for density diagnostics in the range
of typical values of $n_e$ in tokamaks.

The goal of the present work is to study the magnetic-dipole transitions
within the ground configurations of the $3d^n$ ions of tungsten using the NIST
EBIT. Previously we reported several EUV lines in Co-, Ca- and K-like ions
\cite{12286EL,12366EL} within configurations $3d^9$, $3d^2$ and $3d$,
respectively. Here we extend our measurements to include the remaining ions of
tungsten with open $3d$ shell. Using detailed collisional-radiative (CR)
modeling, we identify the measured spectral lines in the EUV range of spectra
between 10 nm and 25 nm. In addition, we perform CR simulations to explore
potential use of the newly identified lines for diagnostics of hot fusion
plasmas.

The paper is organized as follows. Section II describes the experiment and
measurement of the spectra. Details of the CR modeling are presented in
Section III. We then discuss the identification of the new M1 lines. Section V
presents the analysis of the line ratios that can be used for density
diagnostics in fusion plasmas. Finally, the last Section summarizes our
conclusions.

\section{Experiment}

The measurements of spectra from the $3d^n$ ions of tungsten were performed at
the NIST EBIT facility \cite{Gillaspy_1997} using a grazing-incidence EUV
spectrometer \cite{Blagojevic_2005}. The photons were collected by a spherical
gold-coated mirror. A 1:1 image of the EBIT plasma column was focused onto the
spectrometer entrance slit. The mirror center was at an equal distance of
480~mm  from the EBIT axis and the spectrometer entrance slits. The photons
were dispersed with a reflection flat-field grating with 1200 lines/mm. The
grazing incidence angle for both the mirror and the grating was 3$^{\circ}$.
The slit width was kept at 500~$\mu\text{m}$ resulting a constant resolving
power of about 350. The EUV spectra were directly recorded by a liquid
nitrogen cooled back-illuminated charge-coupled device (CCD) that was placed
in the focal plane of the grating at a distance of 235 mm. The CCD detector
has an array of 1340$\times$400 pixels
(20$\mu\text{m}\times\text{20}\mu\text{m}$ each). A detailed description of
our EUV spectroscopic system can be found in \cite{Blagojevic_2005}.

The spectra were measured in two separate runs, one in 2008 (run A) and
another in 2010 (run B). The nominal electron beam energies in run A were 4500
eV, 4750 eV, 5000 eV, 5250 eV, 5500 eV, 6000 eV, and 7000 eV, and the observed
spectra were in the range between 4.5~nm and 19.5~nm. A theoretical analysis
of the spectra indicated that some additional lines may have longer
wavelengths and therefore the second run of measurements was initiated. The
beam energies for run B were selected to be complementary to those for run A,
namely, 4665 eV, 4840 eV, 5155 eV, 5355 eV, 5755 eV, and 6500 eV, and the
observed spectral window was shifted to 8~nm to 26~nm by translating the
detector in the focal plane. The electron beam current for both runs was
150~mA and the trap depth was approximately 220~V. The trap was emptied and
reloaded every 11~s. 

The measured spectra of tungsten were calibrated with lines from lighter
elements. Reference spectra of Ne, Ar, O, and Fe were measured at several
energies between 2 keV and 9 keV for run A. The calibration of spectra for run
B was performed with lines from N, O, Fe and Kr at beam energies between 1 keV
and 16 keV. Both gas injection \cite{gas_injection} and metal vapour vacuum
arc ion source (MEVVA) \cite{MEVVA} systems were utilized in the calibration
runs. The observed calibration lines were fitted with the
statistically-weighted Gaussian line profiles.  The calibration curve was a
fourth-order polynomial fit of the line centers (CCD pixel number) to the
known wavelengths. The weighting in the fit to the calibration curve was based
on the quadrature sum of the statistical uncertainty of our observation of the
calibration line center,  the accuracy of the calibration line wavelength, and
estimated systematic measurement uncertainty. When a wavelength was measured
at various beam energies, the final wavelength was taken to be the weighted
average of the corresponding values (with exceptions noted below), while the
total error in the final wavelength was taken to be the quadrature sum of the
total uncertainty from the calibration curve and the reduced statistical
uncertainty from the average at various energies.  The statistical
uncertainties in the line positions were typically less than 0.001 nm. The
final accuracy of the W spectral lines was 0.003 nm.  

\begin{figure}
\includegraphics[width=1\textwidth]{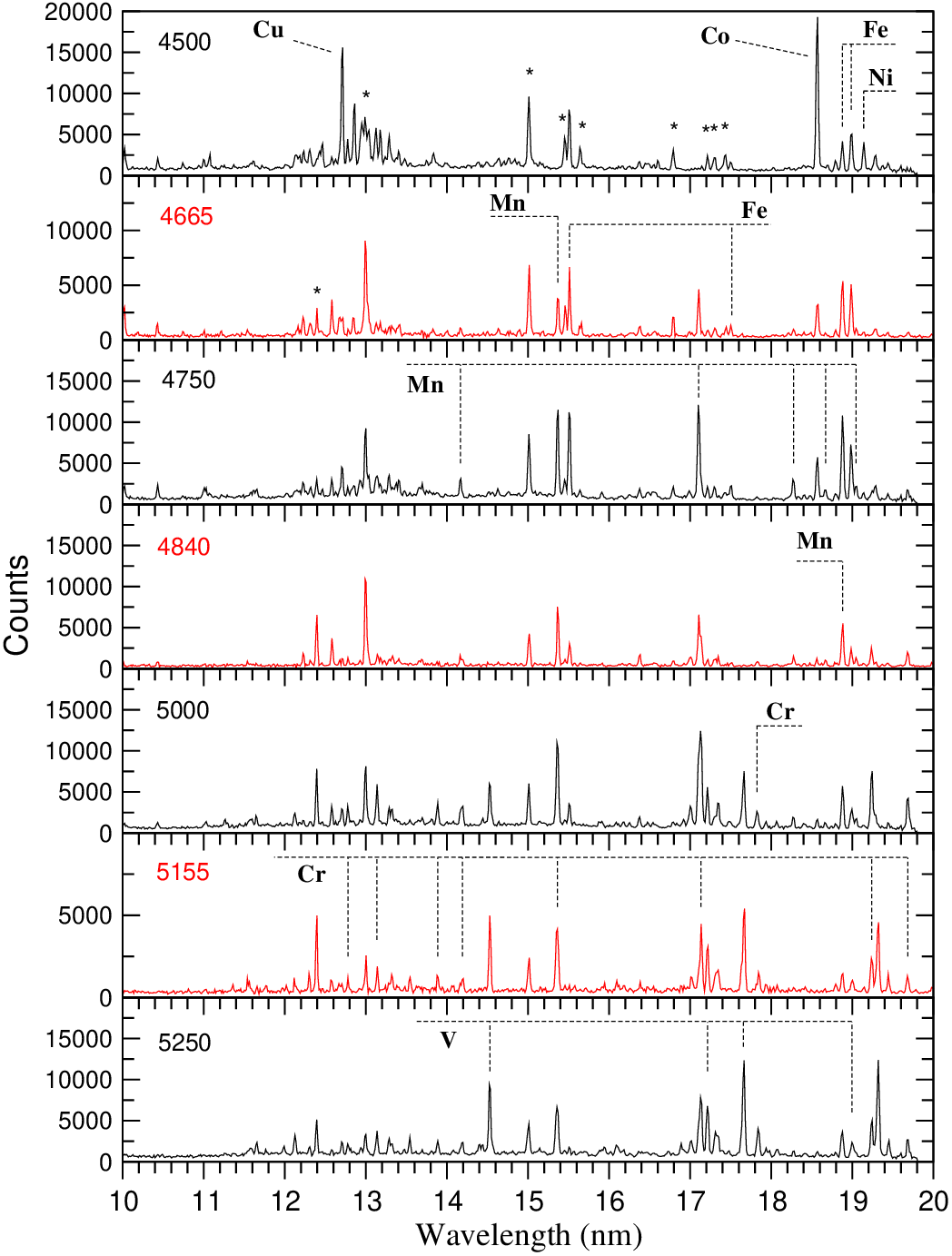}
\caption{\label{Fig1a1} Tungsten spectra between 10 nm and 20 nm for beam
energies between 4500 eV and 5250 eV. The identified transitions are indicated 
by vertical dashed lines. The spectra from run A are shown in black and the spectra from
run B are shown in red. Asterisks show the strongest impurity lines.}
\end{figure}

\begin{figure}
\includegraphics[width=1\textwidth]{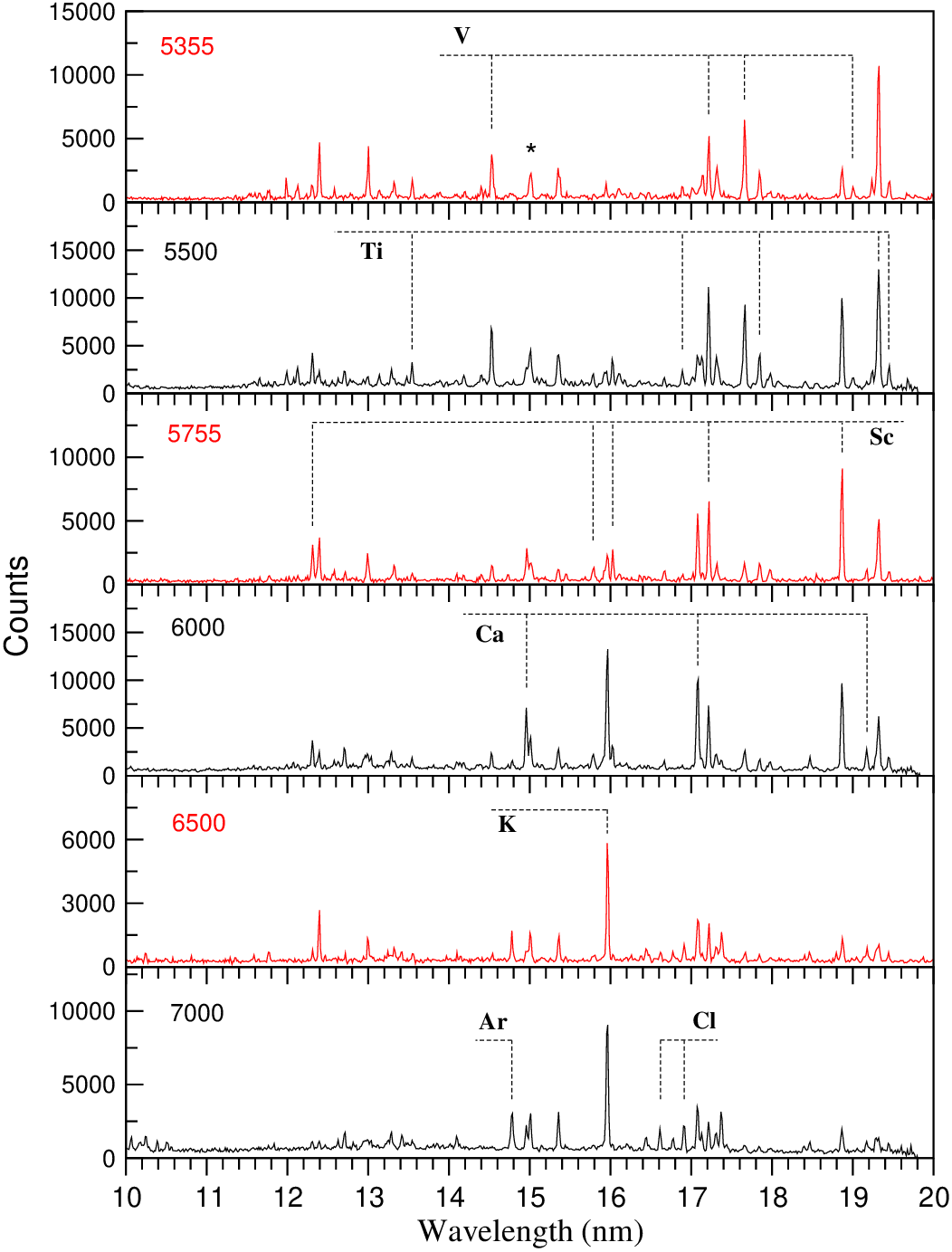}
\caption{\label{Fig1a2} Tungsten spectra between 10 nm and 20 nm for beam
energies between 5355 eV and 7000 eV. The identified transitions are indicated 
by vertical dashed lines. The spectra from run A are shown in black and the spectra from
run B are shown in red. Asterisks show the strongest impurity lines.}
\end{figure}

The measured spectra for tungsten are shown in Figs.~\ref{Fig1a1} (beam
energies $E_B$ = 4500 eV to 5250 eV) and \ref{Fig1a2} (beam energies $E_B$ =
5355 eV to 7000 eV). The run-A spectra are shown in black and the run-B
spectra are presented in red (color online only). The spectral region in the
figures is limited to $\lambda$= 10 nm to 20 nm since almost all M1 lines from
the $3d^n$ ions are within this range. Only one line, $\lambda \approx$
21.203~nm in the V-like ion, was found above 20 nm, and therefore we do not
show the run B spectra at longer wavelengths. The identified transitions in
various ions of tungsten are indicated by vertical dashed lines in the plots.
The measured spectra also contain a few impurity lines from oxygen (e.g., at
15~nm) and xenon, which are marked by asterisks. The highest-energy spectrum of
$E_B$ = 7000 eV also shows a few lines from Ar- and Cl-like ions which have
already been identified in our previous work \cite{12366EL}.

\section{Collisional-radiative modeling of EBIT spectra}

Generally, identification of measured spectral lines greatly benefits from
applying methods that include comparisons of different physical parameters,
such as wavelengths and intensities. While atomic structure methods for simple
ions can calculate wavelengths with the accuracy at the level of 0.01\% or
even better, the simulations for multi-electron ions with open shells may not
be as precise as needed for unambiguous line identification. Another often
used technique in EBIT studies is the analysis of the variation of line
intensity with beam energy. This method, however, would only be of marginal
value when neighboring ions do not differ much in ionization energy, and
therefore it is difficult to uniquely associate a line with a specific
ionization stage.

The most reliable analysis of spectral lines can be accomplished with a
collisional-radiative modeling of EBIT plasmas. For any given set of plasma
parameters, such as beam energy and density, the CR simulations can produce a
detailed synthetic spectrum containing lines from a number of ions. A
comparison of calculated line positions and line intensities with the spectra
measured at several energies provides practically unambiguous identification
of spectral lines. This method was successfully used in our previous
publications \cite{9000EL,8219TP,12286EL,14870EL,12366EL} in order to analyze
and identify dozens of spectral lines from highly-charged heavy ions in x-ray
and EUV regions. 

In this work we implement the non-Maxwellian collisional-radiative code {\sc NOMAD}
\cite{NOMAD} for the calculation of spectra from tungsten ions in the EBIT. The
solution of the steady-state rate equation
\begin{equation}
\label{eq1}
\hat{A} \cdot \hat{N} = 0
\end{equation}
provides populations of all relevant atomic states and, consequently,
intensities of spectral lines. Here $\hat{N}$ is the vector of populations of 
atomic
states included in simulations
and $\hat{A}$ is the rate matrix describing physical processes that affect
state populations. The detailed representation of Eq.~(\ref{eq1})
is:
\begin{eqnarray}
\label{eqrate}
\nonumber \sum_{j>i} {N_{z,j} \cdot \left( A_{z,ij}^{rad}+n_e
R_{z,ij}^{dx}\right)} + \sum_{j<i} {N_{z,j} n_e R_{z,ij}^{ex}} + \sum_{k}{n_e
R_{z-1,ki}^{ion}} + \sum_{k}{n_e R_{z+1,ki}^{rr}} + \delta _{i1} n_0
R_{z+1}^{cx}\\ 
- N_{z,i} \left( \sum_{j<i}{\left( A_{z,ji}^{rad}+n_e
R_{z,ji}^{dx}\right)} + \sum_{j>i} {n_e R_{z,ji}^{ex}} + \sum_{k}{n_e
R_{z,ki}^{ion}} + \sum_{m}{n_e R_{z,mi}^{rr}} + \delta _{i1} n_0
R_{z}^{cx}\right) = 0
\end{eqnarray}
where $N_{z,i}$ is the population of atomic state $j$ in an ion $z$,
$A_{z,ij}^{rad}$ is the radiative transition probability, $n_e$ is the
electron density, $R_{z,ij}^{ex}$, $R_{z,ij}^{dx}$, and $R_{z-1,ki}^{ion}$ are
the rate coefficients for electron-impact excitation, deexcitation and
ionization, respectively, $R_{z,mi}^{rr}$ is the rate coefficient for
radiative recombination, $n_0$ is the density of the neutrals in the trap, and
$R_{z}^{cx}$ is the rate coefficient for the charge exchange (CX) between
neutrals and W ions. Unlike Maxwellian plasmas, dielectronic capture (DC) is
normally neglected in EBIT collisional-radiative simulations since this
resonant process requires an accurate match of the beam energy with the DC
energy. 

Charge exchange between the W ions and neutrals in the ion trap can 
affect the ionization distribution.  The Kronecker factor $\delta _{i1}$ in
Eq.~(\ref{eqrate}) indicates that in our model the CX connects only the ground
states of adjacent ions. It is worth noting that since the ion charge is so
high ($z \approx$ 50), the contribution of double CX may be comparable to the
single CX. We are unaware of any calculations or measurements of single
or multiple CX cross sections between highly-charged  tungsten and neutral
atoms or molecules, and therefore the Classical Trajectory Monte Carlo cross section
scaling~\cite{CTMC} 
\begin{equation}
\sigma_{cx} = z \cdot 10^{-15}~ cm^2 
\label{eqcx}
\end{equation}
was used in calculations. In fact, the precise value of this parameter is
not very important as it enters the rate equations as a factor in the product
$n_0 v_0 \sigma_{cx}$, where $v_0$ is the relative velocity between neutrals
and tungsten ions. Neither $n_0$ nor $v_0$ are accurately known for our experimental
conditions, so that the product $n_0 v_0$ was used as the only free 
parameter in CR simulations. 

The NOMAD code solves the rate equations (\ref{eq1}) using externally
calculated basic atomic data. For the present work the energy levels,
radiative transition probabilities (up to electric and magnetic octupoles) and
electron-impact collisional cross sections were calculated with the
relativistic Flexible Atomic Code (FAC), which is described in detail in
Ref.~\cite{FAC}. The relativistic atomic structure (including
quantum-electrodynamics corrections) and collision methods implemented in FAC
are well suited for highly-charged ions of heavy elements. Our CR model
contains 15 ions, from Zn-like W$^{44+}$ to Si-like W$^{60+}$. Since the ions
with ionization potentials $I_z$ larger than the beam energy $E_b$ have very
small populations, only 6 to 8 ionization stages were kept for each specific
simulation. The atomic states for each ion contained single and double
($\Delta n = 0$ only) excitations from the ground configuration. Single
excitations from the $3l$ subshells were included up to $n$ = 5 for most of
the open-shell ions, and up to $n=7$ or 8 for closed-shell (Ni-like) ions or
ions with one or two electrons above closed shells. The double excitations are
included only for $\Delta n = 0$ within n=3 shell.

In our previous works on high-Z ions with open $s$ and $p$ shells
\cite{9000EL,8219TP,12286EL,14870EL,12366EL} all singly- and doubly-excited
states included in the CR modeling were atomic {\it levels}, i.e., the
fine-structure components. Since open $3d^n$ shells allow many more permitted
combinations of angular momenta, the total number of atomic levels due to
single and double excitations from $3s^23p^63d^n$ increases drastically and
becomes untractable with available computational facilities. While a typical
number of levels in CR modeling of $4s^24p^n$ and $3s^23p^n$ ions was on the
order of 1000 per ionization stage, the excitations from $3s^23p^63d^n$ can
generate 10,000 levels or more, and thus the total number of levels becomes
prohibitively large. 

In order to reduce the size of the rate equations to an acceptable level,  the
atomic states within each $3d^n$ ion were divided into two groups. The first
group was composed of the ground configuration levels $3s^23p^63d^n$ and
singly- and doubly-excited levels within the same n=3 shell, i.e.,
$3s^23p^53d^{n+1}$, $3s3p^63d^{n+1}$, $3s^23p^43d^{n+2}$, $3s3p^53d^{n+2}$,
and $3p^63d^{n+2}$. The levels in this first group were retained without
modification as the fine-structure components. The levels in the second group,
i.e., $\Delta n \geq 1$ excitations $3s^23p^63d^{n-1}kl$, $3s^23p^53d^nkl$ and
$3s3p^63d^nkl$ with $k \geq 4$, were joined into generalized atomic states,
which are referred to as the ``superterms" below. The procedure of level
grouping can be exemplified for the $3d^{n-1}kl$ configuration. Each of the
atomic levels within this configuration can be described by the following set
of quantum numbers in jj-coupling (FAC level notations are given in this
coupling scheme):
$(((3d_{-}^a)_{j_{-}},(3d_{+}^b)_{j_{+}})_{j_c},(kl)_{j_k})_J$ or simply
$((j_{-},j_{+})_{j_c},j_k)_J$ where $a+b=n-1$, $j_{-}$ and $j_{+}$ are the
momenta of the $3d_{-}$ and $3d_{+}$ sub-shells, $j_c$ is the total angular
momentum of the core $3d^{n-1}$, $j_k=l\pm1/2$ is the momentum of the optical
electron $kl$, and $J$ is the total angular momentum. Here and below we use
notation $l_{\pm}$ for an $l$ electron with $j=l\pm 1/2$. The simplest
procedure in level grouping would be to join them according to the atomic
jj-terms $((j_{-},j_{+})_{j_c},j_k)$. However, the reduction in the total
number of states is rather small. Even the next level of grouping, based on
the core momentum $j_c$, results in several thousands of states per ion.
Therefore the excited levels in the present work were joined according to the
$(j_{-},j_{+})$ pairs. For the high excited configurations with a hole in the
$3s$ or $3p$ subshell, the levels were combined by the three momenta
$(j_h,j_{-},j_{+})$ where $j_h$ is the hole momentum. We found that while such
grouping significantly reduces the total number of states per ion, the
resulting set of superterms provides a sufficiently dense representation of
atomic structure for each of the $3d^n$ ions in our CR model.  The actual
reduction in the number of states in an ion can reach an order of magnitude:
for instance, for the V-like ion with $3d^5$ ground configuration the
originally generated 10801 levels are reduced to 791 fine-structure levels and
465 superterms only. This method of including lowest atomic levels and
highly-excited generalized states is similar to the recently proposed hybrid
CR models \cite{HYBRID} for high-density plasma kinetics, where the
highly-excited levels are combined into even more general groups of states,
namely, configurations or superconfigurations.

Another feature of our calculations is the additional correction of calculated
energies for the $3d^n$ levels. For each of these ions we performed another
calculation with the FAC code including, in addition to configurations
mentioned above, all possible excitations within the n=3 complex. (Obviously,
the total complex was already included for $3d^9$ and $3d^8$ ions.) The
energies of the ground state configurations in the CR model were then replaced
by the newly calculated energies which, although different by a fraction of a
percent only, still improved the agreement with the experimental energies and
wavelengths. A similar procedure was applied in our recent work on EUV spectra
from highly-charged ions of Hf, Ta and Au \cite{DragJPB}.

The transition probabilities and cross sections between the levels and the
superterms or between the superterms were derived from the FAC results for
transitions between atomic levels using statistical averaging. The collisional
cross sections were then convolved with a 45-eV-wide Gaussian electron-energy
distribution function of the EBIT beam in order to generate the rate
coefficients. The final set of rate equations (\ref{eq1}) was solved in the
steady-state approximation  for a grid of beam energies between 4300 eV and
7000 eV. For each  energy the charge-exchange parameter $n_0 v_0$ varied
between $10^{13}$ and $3\cdot 10^{14}$ cm$^{-2}$s$^{-1}$. The ionization
distributions, level populations, and spectral line intensities were
calculated for each combination of $E_b$ and $n_0 v_0$, and the spectral
patterns were compared with the measured spectra to find the best agreement.
The best value of $n_0 v_0$ was found to be about 10$^{14}$ cm$^{-2}$s$^{-1}$.
This agrees with our order-of-magnitude estimates of 10$^7$ cm$^{-3}$ for the
density of neutrals and 10$^7$ cm/s for the relative velocity.

\begin{figure}
\includegraphics[width=1\textwidth]{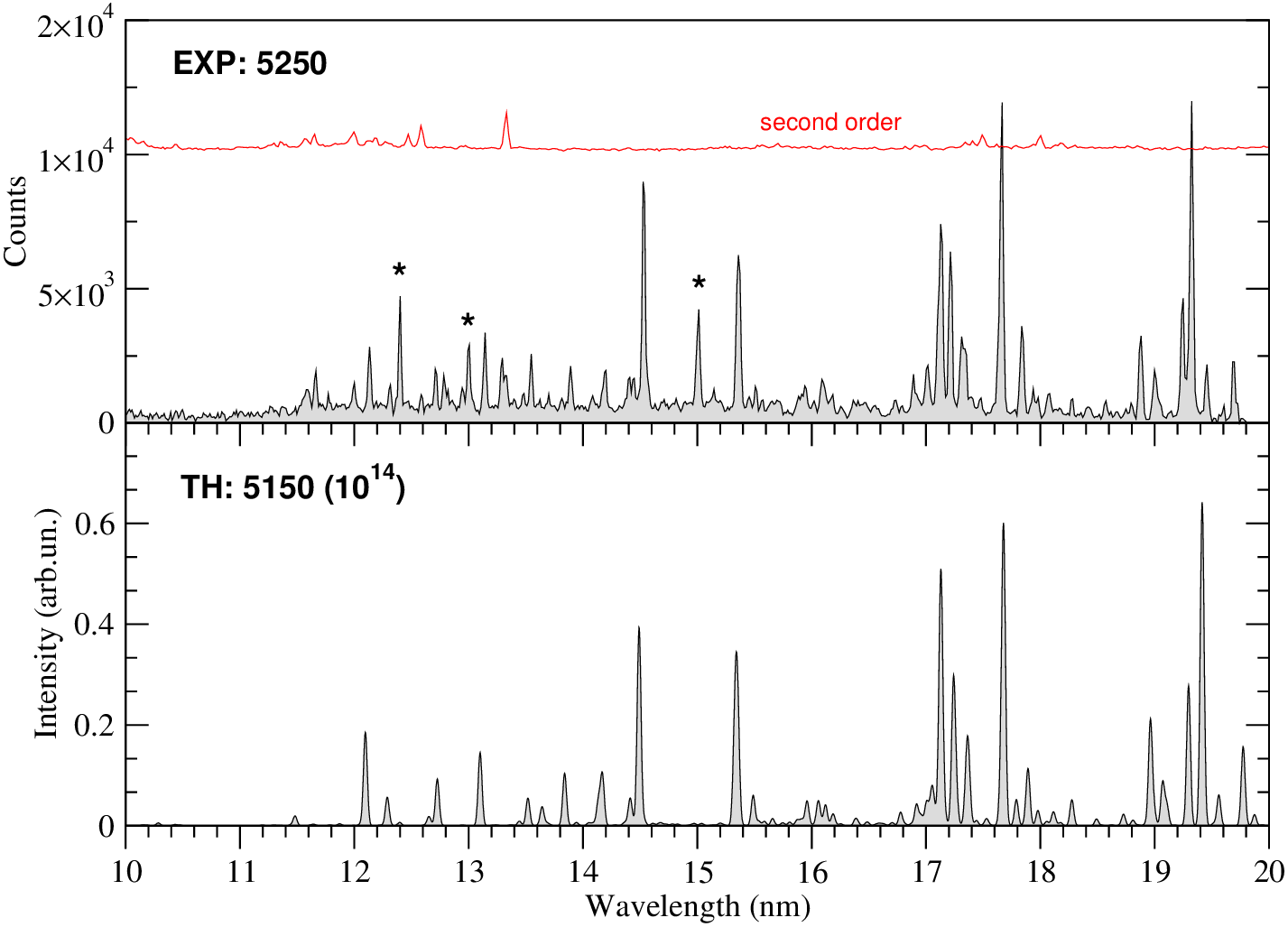}
\caption{\label{FigTH} Comparison of experimental spectrum at the nominal beam
energy of 5250 eV (top) and calculated spectrum at 5150 eV and
$n_0v_0$=10$^{14}$ cm$^{-2}$s$^{-1}$. The second order spectrum is shown by
the shifted line and the strongest impurity lines from Xe (12.4 nm and 13.0 nm)
and O (15.0 nm) are indicated by asterisks.}
\end{figure}

An example of comparison between the experimental and calculated spectra is
presented in Fig.~\ref{FigTH}. The simulated spectrum for the beam energy of
5150 eV with $n_0 v_0=$ $10^{14}$ cm$^{-2}$s$^{-1}$ and CX cross section from
Eq.~(\ref{eqcx}) (bottom) agrees very well with the measured spectral pattern
at the nominal beam energy of 5250 eV (top); this energy difference is
attributed to the space charge effects in the trap. The three strong lines at
12.4~nm, 13.0~nm, and 15.0~nm marked by asterisks are due to  xenon and
oxygen impurities.  Also, Fig.~\ref{FigTH} (top) shows the second order
spectrum shifted along the vertical axis in order to indicate a few relatively
weak second-order lines. One can see that both line positions and line
intensities are reproduced in our simulations very accurately so that most of
the lines can be identified from the visual comparison. Such a good agreement
was observed for all cases considered in the present work.

\section{Line identification}

Table~\ref{Tab1} presents the strongest identified lines between 10 nm and 25
nm in the experimental spectra of runs A and B. Almost all lines in this table
are the forbidden magnetic-dipole transitions within the ground 
configurations $3d^n$ of tungsten ions from Co-like W$^{47+}$ to K-like
W$^{55+}$. The only exception is the 18.468-nm M1 line within the lowest
excited  configuration $3p^53d^2$ of the K-like ion. Four of the observed
lines, namely, 18.567~nm in the Co-like ion, 17.080~nm and 14.959~nm in the
Ca-like ion and 15.962~nm in the K-like ion, are already known from our
previous measurements \cite{12366EL}. The other lines in Table~\ref{Tab1} are
reported here for the first time. As discussed above, the uncertainty of the
measured wavelengths is $\pm$0.003 nm. 

\begingroup
\squeezetable
\begin{longtable*}{cclcllc}
\caption[]{Identified magnetic dipole lines in the experimental
spectra between 10 nm and 25 nm. The previously known lines are marked by
asterisks. The FAC level numbers within ions are given in square brackets. Other theoretical works: a--\cite{11741EL}, b--\cite{7416TP}, c--\cite{8725TP},
d--\cite{7367TP}.}\\
\hline
 Ion   & Sequence & $\lambda_{exp}$ & $\lambda_{th}$ &             &          & A \\
charge &      &(nm)             & (nm)           & Lower level & Upper level &  (s$^{-1}$) \\
\hline
\endhead
\label{Tab1}
47 & Co &  18.567* & 18.640,18.6229$^a$ & $3d^9$ [1] ($d_{+}^5$)$_{5/2}$ &
$3d^9$ [2] ($d_{-}^3$)$_{3/2}$ &
2.47(6) \\
\\
48 & Fe &  15.511 & 15.525 & $3d^8$ [1] ($d_{+}^4$)$_{4}$ & $3d^8$ [6]
(($d_{-}^3$)$_{3/2}$,($d_{+}^5$)$_{5/2}$)$_{4}$ &
1.01(6) \\
48 & Fe &  17.502 & 17.489 & $3d^8$ [2] ($d_{+}^4$)$_{2}$ & $3d^8$ [7]
(($d_{-}^3$)$_{3/2}$,($d_{+}^5$)$_{5/2}$)$_{1}$ &
1.71(6) \\
48 & Fe &  18.878 & 18.956 & $3d^8$ [2] ($d_{+}^4$)$_{2}$ & $3d^8$ [5]
(($d_{-}^3$)$_{3/2}$,($d_{+}^5$)$_{5/2}$)$_{2}$ &
1.93(6) \\
48 & Fe &  18.988 & 19.075 & $3d^8$ [1] ($d_{+}^4$)$_{4}$ & $3d^8$ [4]
(($d_{-}^3$)$_{3/2}$,($d_{+}^5$)$_{5/2}$)$_{3}$ &
3.22(6) \\
\\
49 & Mn &  14.166 & 14.139 & $3d^7$ [1] ($d_{+}^3$)$_{9/2}$ & $3d^7$ [10]
(($d_{-}^3$)$_{3/2}$,($d_{+}^4$)$_{2}$)$_{7/2}$ &
1.76(5) \\
49 & Mn &  15.368 & 15.354 & $3d^7$ [1] ($d_{+}^3$)$_{9/2}$ & $3d^7$ [9]
(($d_{-}^3$)$_{3/2}$,($d_{+}^4$)$_{4}$)$_{11/2}$ &
1.89(5) \\
49 & Mn &  17.106 & 17.137 & $3d^7$ [1] ($d_{+}^3$)$_{9/2}$ & $3d^7$ [5]
(($d_{-}^3$)$_{3/2}$,($d_{+}^4$)$_{2}$)$_{9/2}$ &
2.23(6) \\
49 & Mn &  18.276 & 18.303 & $3d^7$ [3] ($d_{+}^3$)$_{5/2}$ & $3d^7$ [10]
(($d_{-}^3$)$_{3/2}$,($d_{+}^4$)$_{2}$)$_{7/2}$ &
2.72(5) \\
49 & Mn &  18.670 & 18.741 & $3d^7$ [2] ($d_{+}^3$)$_{3/2}$ & $3d^7$ [8]
(($d_{-}^3$)$_{3/2}$,($d_{+}^4$)$_{2}$)$_{1/2}$ &
2.56(6) \\
49 & Mn &  18.880 & 18.972 & $3d^7$ [1] ($d_{+}^3$)$_{9/2}$ & $3d^7$ [4]
(($d_{-}^3$)$_{3/2}$,($d_{+}^4$)$_{2}$)$_{7/2}$ &
3.57(6) \\
49 & Mn & 19.047 & 19.127 & $3d^7$ [2] ($d_{+}^3$)$_{3/2}$ & $3d^7$ [7]
(($d_{-}^3$)$_{3/2}$,($d_{+}^4$)$_{4}$)$_{5/2}$ &
1.03(6) \\
\\
50 & Cr &  12.779 & 12.685 & $3d^6$ [1] ($d_{+}^2$)$_{4}$ & $3d^6$ [15]
(($d_{-}^3$)$_{3/2}$,($d_{+}^3$)$_{5/2}$)$_{3}$ &
4.05(5) \\
50 & Cr &  13.137 & 13.105 & $3d^6$ [1] ($d_{+}^2$)$_{4}$ & $3d^6$ [12]
(($d_{-}^3$)$_{3/2}$,($d_{+}^3$)$_{5/2}$)$_{4}$ &
3.68(4) \\
50 & Cr &  13.886 & 13.848 & $3d^6$ [2] ($d_{+}^2$)$_{2}$ & $3d^6$ [15]
(($d_{-}^3$)$_{3/2}$,($d_{+}^3$)$_{5/2}$)$_{3}$ &
4.92(5) \\
50 & Cr &  14.193 & 14.176 & $3d^6$ [2] ($d_{+}^2$)$_{2}$ & $3d^6$ [13]
(($d_{-}^3$)$_{3/2}$,($d_{+}^3$)$_{5/2}$)$_{2}$ &
3.20(5) \\
50 & Cr &  15.363 & 15.363 & $3d^6$ [1] ($d_{+}^2$)$_{4}$ & $3d^6$ [10]
(($d_{-}^3$)$_{3/2}$,($d_{+}^3$)$_{3/2}$)$_{3}$ &
1.01(6) \\
50 & Cr &  17.133 & 17.153 & $3d^6$ [1] ($d_{+}^2$)$_{4}$ & $3d^6$ [8]
(($d_{-}^3$)$_{3/2}$,($d_{+}^3$)$_{9/2}$)$_{5}$ &
6.57(5) \\
50 & Cr &  17.826 & 17.823 & $3d^6$ [3] ($d_{+}^2$)$_{0}$ & $3d^6$ [14]
(($d_{-}^3$)$_{3/2}$,($d_{+}^3$)$_{5/2}$)$_{1}$ &
1.32(6) \\
50 & Cr &  19.239 & 19.317 & $3d^6$ [1] ($d_{+}^2$)$_{4}$ & $3d^6$ [5]
(($d_{-}^3$)$_{3/2}$,($d_{+}^3$)$_{9/2}$)$_{4}$ &
3.02(6) \\
50 & Cr &  19.684 & 19.791 & $3d^6$ [1] ($d_{+}^2$)$_{4}$ & $3d^6$ [4]
(($d_{-}^3$)$_{3/2}$,($d_{+}^3$)$_{9/2}$)$_{3}$ &
2.56(6) \\
\\
51 & V  &  14.531 & 14.511 & $3d^5$ [1] ($d_{+}$)$_{5/2}$ & $3d^5$ [9]
(($d_{-}^3$)$_{3/2}$,($d_{+}^2$)$_{2}$)$_{7/2}$ &
1.21(5) \\
51 & V  &  17.215 & 17.260 & $3d^5$ [1] ($d_{+}$)$_{5/2}$ & $3d^5$ [5]
(($d_{-}^3$)$_{3/2}$,($d_{+}^2$)$_{2}$)$_{3/2}$ &
3.75(6) \\
51 & V  &  17.660 & 17.709 & $3d^5$ [1] ($d_{+}$)$_{5/2}$ & $3d^5$ [3]
(($d_{-}^3$)$_{3/2}$,($d_{+}^2$)$_{4}$)$_{7/2}$ &
1.59(6) \\
51 & V  &  18.996 & 19.098 & $3d^5$ [4] (($d_{-}^3$)$_{3/2}$,($d_{+}^2$)$_{4}$)$_{11/2}$  &
$3d^5$ [13]
(($d_{-}^2$)$_{3}$,($d_{+}^3$)$_{9/2}$)$_{11/2}$ &
2.31(6) \\
51 & V  &  21.203 & 21.370 & $3d^5$ [1] ($d_{+}$)$_{5/2}$ & $3d^5$ [2]
(($d_{-}^3$)$_{3/2}$,($d_{+}^2$)$_{4}$)$_{5/2}$ &
3.40(6) \\
\\
52 & Ti &  13.543 & 13.521 & $3d^4$ [5] (($d_{-}^3$)$_{3/2}$,($d_{+}$)$_{5/2}$)$_{3}$  &
$3d^4$ [17]
(($d_{-}^2$)$_{0}$,($d_{+}^2$)$_{4}$)$_{4}$ &
1.09(6) \\
52 & Ti &  16.890 & 16.922 & $3d^4$ [2] (($d_{-}^3$)$_{3/2}$,($d_{+}$)$_{5/2}$)$_{1}$  &
$3d^4$ [7]
(($d_{-}^2$)$_{2}$,($d_{+}^2$)$_{4}$)$_{2}$ &
4.70(6) \\
52 & Ti & 17.846 & 17.905 & $3d^4$ [3]
(($d_{-}^3$)$_{3/2}$,($d_{+}$)$_{5/2}$)$_{4}$ & $3d^4$ [10]
(($d_{-}^2$)$_{2}$,($d_{+}^2$)$_{4}$)$_{5}$ &
1.65(6) \\
52 & Ti &  19.319 & 19.427,19.6$^b$ & $3d^4$ [1] ($d_{-}^4$)$_{0}$ & $3d^4$ [2]
(($d_{-}^3$)$_{3/2}$,($d_{+}$)$_{5/2}$)$_{1}$ &
3.31(6) \\
52 & Ti &  19.445 & 19.568 & $3d^4$ [3] (($d_{-}^3$)$_{3/2}$,($d_{+}$)$_{5/2}$)$_{4}$  &
$3d^4$ [8]
(($d_{-}^2$)$_{2}$,($d_{+}^2$)$_{4}$)$_{4}$ &
3.02(6) \\
\\
53 & Sc &  12.312 & 12.291 & $3d^3$ [1] ($d_{-}^3$)$_{3/2}$ & $3d^3$ [7]
(($d_{-}^2$)$_{0}$,($d_{+}$)$_{5/2}$)$_{5/2}$ &
2.75(5) \\
53 & Sc &  15.785 & 15.812 & $3d^3$ [4] (($d_{-}^2$)$_{2}$,($d_{+}$)$_{5/2}$)$_{9/2}$  &
$3d^3$ [12]
(($d_{-}$)$_{3/2}$,($d_{+}^2$)$_{4}$)$_{11/2}$ &
1.42(6) \\
53 & Sc &  16.027 & 16.056 & $3d^3$ [1] ($d_{-}^3$)$_{3/2}$ & $3d^3$ [6]
(($d_{-}^2$)$_{2}$,($d_{+}$)$_{5/2}$)$_{1/2}$ &
1.02(6) \\
53 & Sc & 17.216 & 17.271 & $3d^3$ [1] ($d_{-}^3$)$_{3/2}$ & $3d^3$ [3]
(($d_{-}^2$)$_{2}$,($d_{+}$)$_{5/2}$)$_{3/2}$ &
2.74(6) \\
53 & Sc &  18.867 & 18.971 & $3d^3$ [1] ($d_{-}^3$)$_{3/2}$ & $3d^3$ [2]
(($d_{-}^2$)$_{2}$,($d_{+}$)$_{5/2}$)$_{5/2}$ &
3.41(6) \\
\\
54 & Ca &  14.959* & 14.984,15.010$^c$ & $3d^2$ [1] ($d_{-}^2$)$_{2}$ & $3d^2$ [4]
(($d_{-}$)$_{3/2}$,($d_{+}$)$_{5/2}$)$_{2}$ &
1.81(6),1.798(6)$^c$ \\
54 & Ca &  17.080* & 17.147,17.157$^c$ & $3d^2$ [1] ($d_{-}^2$)$_{2}$ & $3d^2$ [3]
(($d_{-}$)$_{3/2}$,($d_{+}$)$_{5/2}$)$_{3}$ &
3.68(6),3.683(6)$^c$ \\
54 & Ca &  19.177 & 19.281,19.294$^c$ & $3d^2$ [2] ($d_{-}^2$)$_{0}$ & $3d^2$ [6]
(($d_{-}$)$_{3/2}$,($d_{+}$)$_{5/2}$)$_{1}$ &
1.72(6),1.771(6)$^c$ \\
\\
55 & K  &  15.962* & 16.003 & $3d$ [1] ($d_{-}$)$_{3/2}$ & $3d$ [2]
($d_{+}$)$_{5/2}$&
2.59(6),1.48(6)$^d$ \\
55 & K  &  18.468 & 18.536 & $3p^53d^2$ [6]
(($p_{+}^3$)$_{3/2}$,($d_{-}^2$)$_{2}$)$_{7/2}$ & $3p^53d^2$ [9]
((($p_{+}^3$)$_{3/2}$,$d_{-}$)$_{3}$,$d_{+}$)$_{9/2}$  &
2.99(6) \\
\hline
\end{longtable*}
\endgroup

The atomic levels in Table~\ref{Tab1} are described in jj-coupling, as
calculated by the FAC code. The $l_{\pm}$ groups with total zero angular
momentum are not shown. For instance, the excited level of the $3d^9$
configuration of the Co-like ion has six $3d_{+}$ electrons with momentum
projections from $m_j$ = -5/2 to $m_j$ = +5/2 which are omitted in the
notation. The numbers in square brackets in the level notation columns show
the calculated level number within the corresponding ion (the ground level is
number 1 and so on).

There are several lines from neighboring ions that have very close
wavelengths, e.g., 18.878~nm in Fe-like and 18.880~nm in Mn-like ion, or
15.368~nm in Mn-like and 15.363~nm in Cr-like ion. For such cases the
wavelengths were determined from a spectrum where one of the lines was strong
while the other was weak due to the shifted ionization distribution. The
wavelengths for other lines were obtained by averaging over several measured
spectra.

Table~\ref{Tab1} also shows our calculated wavelengths and transition
probabilities as well as several other theoretical results
\cite{11741EL,7416TP,7367TP,8725TP}. In most cases the present results agree
quite well with the measured wavelengths although for several lines the
difference is rather large, as much as 0.5~\%. This probably reflects
difficulties in atomic structure calculations for such complex ions. Most of
the calculated transition probabilities are between $10^5$ s$^{-1}$ and $5
\cdot 10^6$ s$^{-1}$. The only line with a smaller probability of A = $3.68
\cdot 10^4$ s$^{-1}$ is the 13.137 nm J=4--J=4 transition in Cr-like W$^{}$.
Note also that the recent RMBPT calculations for Ca-like W \cite{8725TP} agree
with our transition probabilities to within a few percent.

\begin{figure}
\includegraphics[width=1\textwidth]{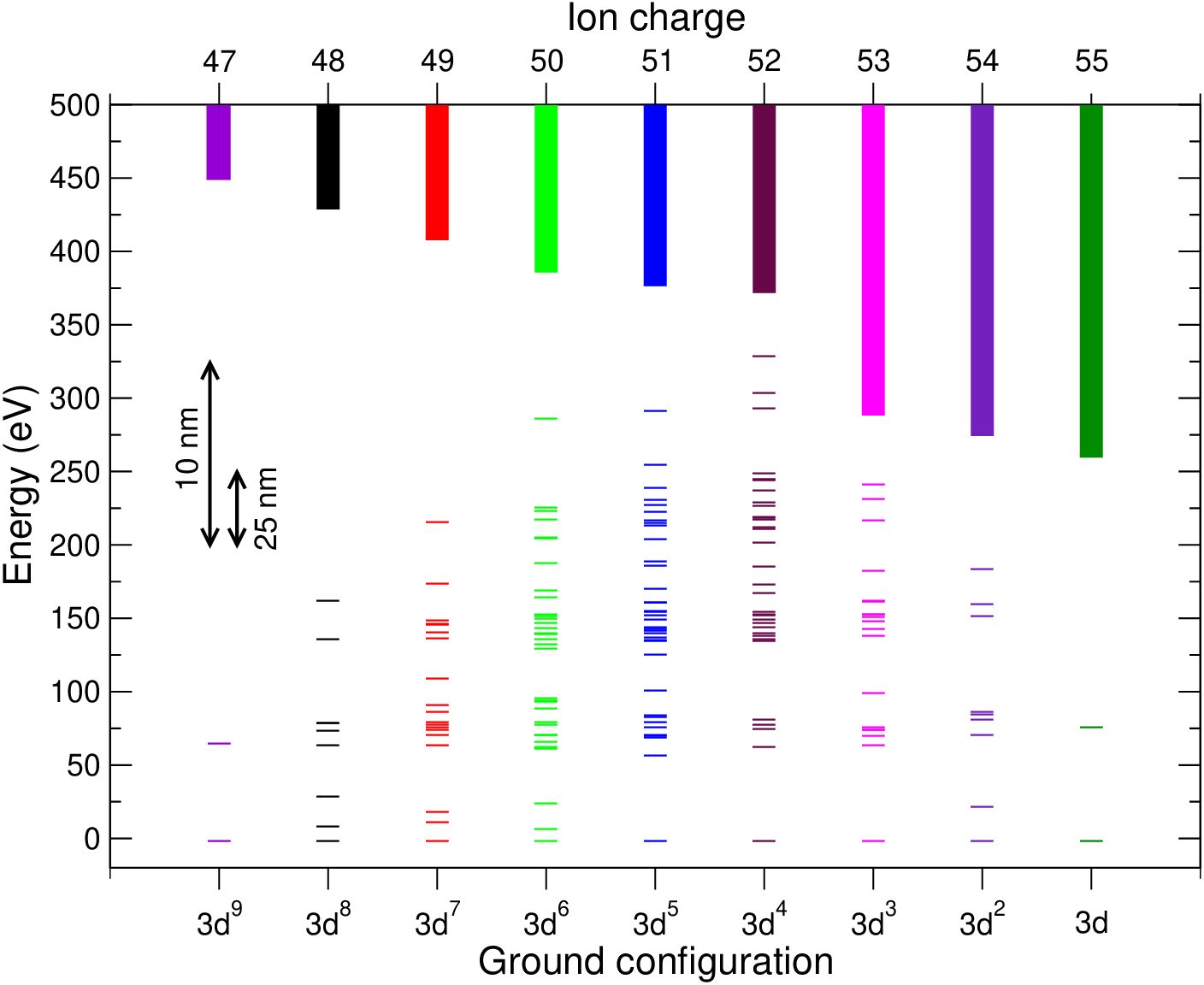}
\caption{\label{FigLevs} Calculated energy levels of the $3p^63d^n$ configurations in W$^{47+}$
through W$^{55+}$. The vertical bars at the top show the spread of the
$3p^53d^{n+1}$ configurations below 500 eV.}
\end{figure}

The energy structure of the $3d^n$ ions and population flux analysis  explain
why only the forbidden M1 lines are visible between 10 nm ($\Delta E \approx$
124 eV) and 25 nm ($\Delta E \approx$ 50 eV) under EBIT conditions. Normally,
the strongest E1 lines in a collisionally-dominated spectrum are due to
transitions between the ground configuration and lowest excited configuration
of opposite parity. There is, however, a relatively large energy gap between
$3p^63d^n$ and $3p^53d^{n+1}$. Our calculations with FAC show that the $3p-3d$
excitation energy in tungsten ions is about (300--400) eV.
Figure~\ref{FigLevs} shows the calculated energy levels below 500 eV for
Co-like through K-like ions of W. The levels of the ground configuration are
represented by horizontal lines, and the vertical bars show the spread of the
$3p^53d^{n+1}$ configuration. For W$^{47+}$, W$^{48+}$,W$^{49+}$, and
W$^{55+}$, the energy gap between these configurations is larger than 124 eV,
so that the corresponding E1 lines have wavelengths smaller than 10 nm. For
the remaining ions, the highest $3d^n$ levels are rather close to the lower
edge of $3p^53d^{n+1}$ manifold. However, only a few possible EUV transitions
obey the $|\Delta J| \leq 1$ selection rule for E1 lines and moreover, those
transitions are greatly suppressed by small branching ratios due to stronger
soft x-ray decays into the lowest levels of the ground configuration.

Some transitions between higher-$n$ states, for instance, n=4--5 transitions
in Fe-like and Mn-like ions, also fall into the 10--25 nm range. In
low-density plasmas, the populations of the lowest excited levels of $3d^n$
are much higher that those of the high-excited states and therefore only M1
lines are strong. When the density is high, the populations approach the
Boltzmann values which are of the same order for all levels in an ion. Since
E1 rates between the higher states are much stronger than the M1
probabilities, only the E1 lines will be present at high densities.

As mentioned above, four of the spectral lines in table~\ref{Tab1} have
already been observed in our previous experiments. While the new wavelengths
for the lines from Ca- and K-like ions agree with the known values within
experimental uncertainties, the 18.567$\pm$0.003 nm wavelength for the M1
transition in Co-like ion is shifted with respect to our previous value of
18.578$\pm$0.002 nm \cite{8219TP}. To address this problem, we reexamined the
4228 eV spectrum of Ref.~\cite{8219TP} where this line was identified for the
first time. As was pointed out in the original publication, the line was
strongly blended by third-order lines; our current analysis shows that its
wavelength should therefore have been assigned a larger uncertainty. For the
lowest beam energies of the present experiment, the M1 line in the Co-like ion
is the strongest in the spectrum (Fig.~\ref{Fig1a1}) so that its wavelength
was determined very reliably. Therefore, the presently measured wavelength of
18.567$\pm$0.003 nm replaces the previous value of Ref.~\cite{8219TP}. Our new
wavelength agrees better with the semi-empirical wavelength of
18.541$\pm$0.032 nm \cite{7017EL}.

\section{Diagnostics with the M1 lines}

The diagnostic potential of the M1 lines is based on several features. First,
the intensity ratios for lines from different ions can conveniently serve as a
diagnostics of temperature and ionization balance. It is also helpful that the
spectral window for these lines is rather narrow, which reduces dependence of
spectrometer efficiency on wavelength. Finally, and most importantly, these
forbidden lines can be used to diagnose electron density in fusion plasmas.

Various methods have been developed for spectroscopic diagnostics of electron
density \cite{GriemBook}. Such techniques make use, for instance, of line
intensity ratios or collisional widths of isolated lines. The line ratio
diagnostics is normally based on comparison of allowed and forbidden lines
which are populated by similar mechanisms (normally, by excitation from the
ground state) and whose radiative decay rates differ by orders of magnitude.
For low densities, when collisional depopulation rates are much smaller than
either of the radiative rates, the intensities of both allowed and forbidden
lines vary linearly with density and therefore their ratio is independent of
$n_e$. When collisional rates become comparable with the probabilities of
forbidden transitions, the ratio shows sensitivity to $n_e$, typically over
one or two orders in $n_e$. A well-known example is the
resonance-to-intercombination-line ratio in He-like ions which has been widely
used in plasma diagnostics \cite{Kunze}.

While the measured low-density spectra from the $3d^n$ ions of tungsten
contain no strong E1 lines, the decay rates for the observed M1 lines vary by
as much as two orders of magnitude, and hence one may expect at least some
sensitivity of line ratios to density variations. In order to analyze the
$n_e$-dependence of the M1 line intensities under typical conditions of hot
fusion plasmas, we performed another set of calculations with NOMAD using a
Maxwellian electron energy distribution function and including dielectronic
recombination within the Burgess-Merts-Cowan-Magee approximation \cite{BMCM}.
The steady-state solutions of the rate equations were determined for electron
densities in the range of $n_e$ = $10^{10}-10^{17}$ cm$^{-3}$ at electron
temperatures $T_e^z \approx I_z$ where $I_z$ is the ionization potential of
the ion under study. It was recently shown \cite{Wiondist} that unlike the
low-Z elements, the typical temperatures of the maximal abundance for
highly-charged W ions are on the order or even larger than the corresponding
ionization potentials, and therefore the condition $T_e^z \approx I_z$ is well
justified for hot steady-state plasmas. Also, since the excitation energies of
the levels within ground configurations are much smaller than $T_e^z$, the
intensity ratios for the EUV lines would only be marginally sensitive to
electron temperature. Finally, since the intensity ratios involve only lines
from the same ionization stage, the conclusions will not depend on ionization
distribution.

\begin{figure}
\includegraphics[width=1\textwidth]{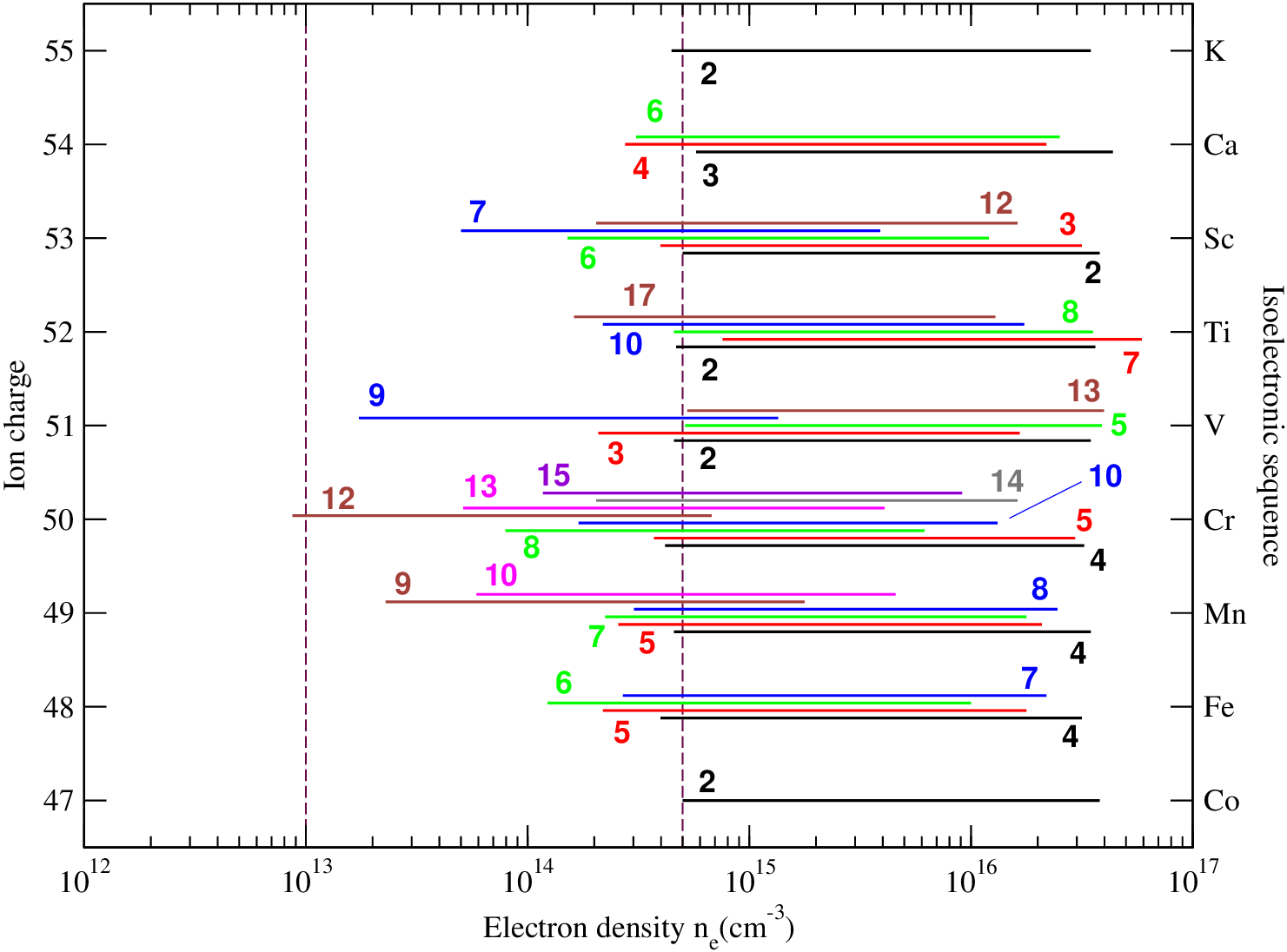}
\caption{\label{DensLim} Transitional ($0.1 < \alpha_i < 0.9$) regions of
densities for the upper levels of M1 lines from table~\ref{Tab1}. The
numbers are the calculated level numbers in a corresponding ion. The range of
typical electron densities of core fusion plasmas is shown by vertical dashed
lines.}
\end{figure}

We have already mentioned above that the collisional depopulation of the upper level is 
the primary physical process resulting in density sensitivity of a spectral
line. The effect of collisions on level population can be approximately
parameterized by their fraction in the total depopulation rate (see also
\cite{Kunze}):
\begin{equation}
\alpha_i(n_e) = \frac{\sum_{j}{n_e R_{ij}^{col}}}{\sum_{j<i}{A_{ij}^{rad}} + 
\sum_{j}{n_e R_{ij}^{col}}}
\label{eqalpha}
\end{equation}
where $\sum_{}{n_eR_{ij}^{col}}$ includes all collisional processes
depopulating level $i$, such as excitation, deexcitation and ionization. The
low-density coronal limit corresponds to $\alpha_i \rightarrow 0$, and in the
high-density Boltzmann (LTE) limit $\alpha_i \rightarrow 1$. The transitional
region for an atomic state can be defined by the condition 
\begin{equation}
0.1 < \alpha_i(n_e) < 0.9,
\label{trans}
\end{equation}
which covers about two orders of magnitude in $n_e$, as follows from
Eq.~(\ref{eqalpha}). This region of $n_e$ determines the range of densities
for a spectral line that are most promising for diagnostics when compared with
other lines that are still in a coronal limit. Figure~\ref{DensLim} shows the
calculated transitional regions of $n_e$ for the upper levels listed in
table~\ref{Tab1} including, for completeness, the levels of Co-like $3d^9$ and
K-like $3d$ ions. It follows from this plot that, for instance, level 9 in the
Mn-like ion, from which the 15.368~nm line originates, decays only radiatively
($\alpha < 0.1$) at densities smaller that $2 \cdot 10^{13}$~cm$^{-3}$, while
electron collisions are the dominant depletion mechanism for this level above
$1.5 \cdot 10^{15}$~cm$^{-3}$. One can see that for almost every ion there
exist levels with rather different ranges of $\alpha_i$ variations, and
therefore one may expect to find several pairs of lines whose intensity ratio
may be used for density diagnostics. Moreover, the transitional regions for a
large number of levels overlap with the typical range of $n_e$ in core fusion
plasmas (marked by the vertical dashed lines). 

\begin{figure}
\includegraphics[width=1\textwidth]{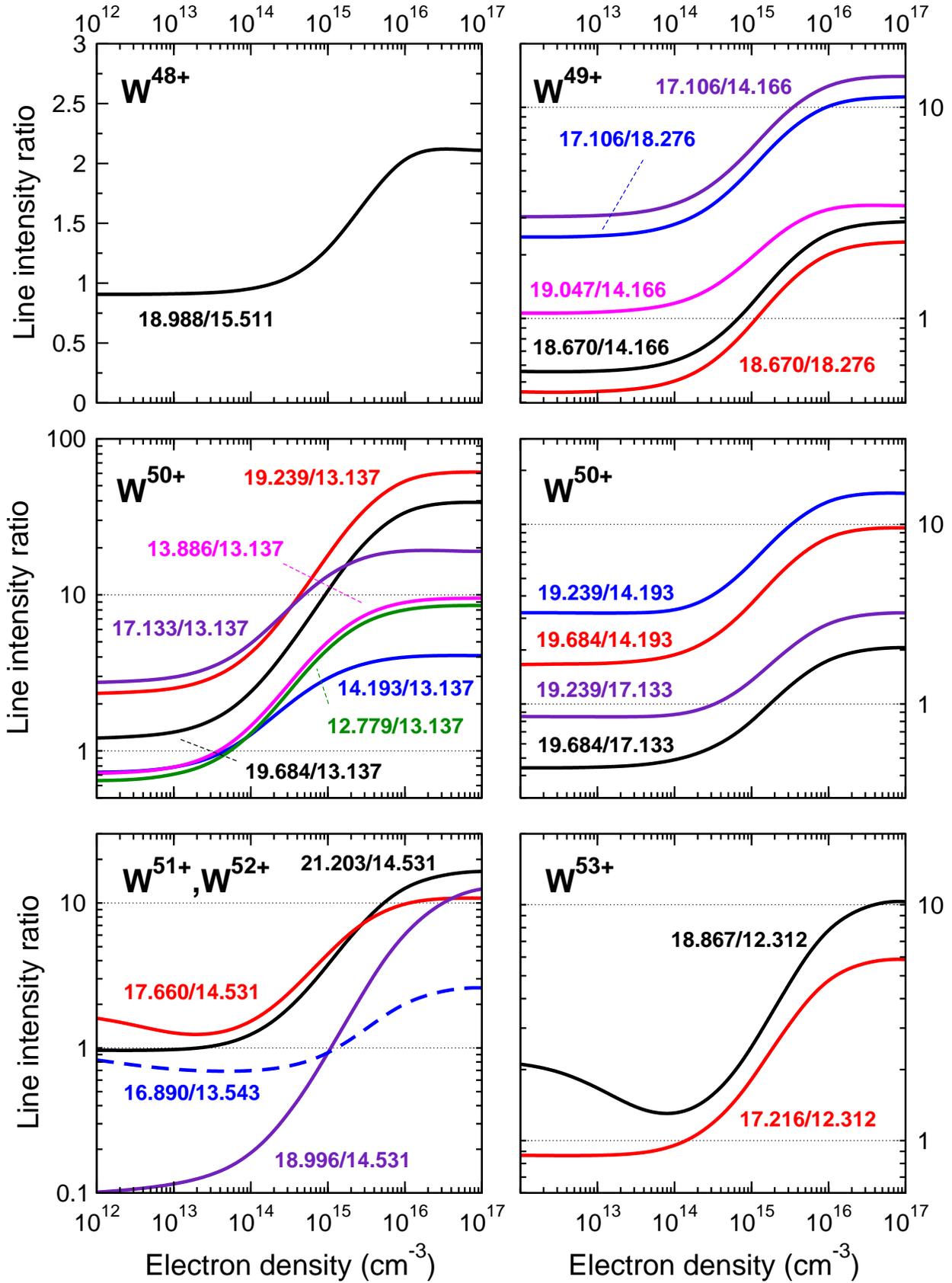}
\caption{\label{Fig6} Density-sensitive line ratios for ions of
tungsten from W$^{48+}$ to W$^{53+}$.}
\end{figure}

The best $n_e$-sensitive line intensity ratios for ions of tungsten from W$^{48+}$ to
W$^{53+}$ are presented in Fig.~\ref{Fig6} and are discussed below. The line
intensities were defined as $I=N \cdot A \cdot \Delta E$ where $N$ is the
upper level population (in cm$^{-3}$), $A$ is the transition probability
(in s$^{-1}$), and $\Delta E$ is the photon energy (in J).

\subsection{Fe-like W$^{48+}$}

Among the four identified lines in the Fe-like ion, the ratio of lines
originating from levels 4 and 6 offers the best sensitivity to density
variations. However, the transitional regions of $n_e$ for these levels, as
follows from Fig.~\ref{DensLim}, are not too different and therefore the line
ratio would not change significantly. Indeed, as shown in the top left panel
of Fig.~\ref{Fig6}, the intensity ratio for the spectral lines at 18.988 nm
and 15.511 nm varies only within a factor of 2.3 between $3 \cdot 10^{14}$
cm$^{-3}$ and 10$^{16}$ cm$^{-3}$. Another complication arises from the
overlap between the 18.988 nm lines and the 19.047 nm line in the next Mn-like
ion W$^{49+}$.

\subsection{Mn-like W$^{49+}$}

Although the seven identified transitions in Mn-like W$^{49+}$ offer several
pairs of lines of potential usage in density diagnostics (top right panel in
Fig.~\ref{Fig6}), it would be rather challenging for experimentalists to
reliably isolate most of these lines from the other lines that originate from
the neighboring ions. It seems that the 18.670/18.276 ratio that increases by
a factor of 5 between $10^{14}$~cm$^{-3}$ and 10$^{16}$~cm$^{-3}$ may actually
be the best choice since these two lines are well isolated in the measured
spectrum.

\subsection{Cr-like W$^{50+}$}

Among all $3d^n$ ions of tungsten, the Cr-like ion offers the largest number
of $n_e$-sensitive line pairs. This is due to both the largest number of
identified lines and the most separated regions of density sensitivity for
different lines. The population of level 12, which is responsible for the
strong and well isolated line at 13.137~nm, starts to deviate from coronal
behavior at electron densities as low as $9 \cdot 10^{12}$~cm$^{-3}$
(Fig.~\ref{DensLim}). The strongest radiative transition from this level has
probability of only $3.68 \cdot 10^4$ s$^{-1}$ as compared with the typical M1
probabilities of 10$^5$~s$^{-1}$ and larger (Table~\ref{Tab1}) and therefore
this level becomes collisionally depleted at lower electron densities. The
line ratios involving the 13.137~nm line, presented in the middle left panel
of Fig.~\ref{Fig6}, show the strongest dependence on $n_e$: for instance, the
19.684/13.137 ratio of two strong and isolated lines varies by a factor of 30
between 10$^{12}$~cm$^{-3}$ and 10$^{16}$~cm$^{-3}$. The ratios with other
lines (middle right panel) vary within smaller limits, from 3 to 6.

\subsection{V-like W$^{51+}$}

The transitional regions for the levels of the V-like ion are mostly between
$2 \cdot 10^{14}$ cm$^{-3}$ and $2 \cdot 10^{16}$ cm$^{-3}$. However, electron
collisions start to depopulate level 9 at much lower densities of about $2
\cdot 10^{13}$ cm$^{-3}$ and thus the line ratios involving a strong isolated
line at 14.531 nm can be very sensitive to electron density. For instance, the
ratio 18.996/14.531 varies by two orders of magnitude, between 0.1 and 10, for
the electron density range of (10$^{13}$--10$^{17}$) cm$^{-3}$.

\subsection{Ti-like W$^{52+}$}

Figure~\ref{DensLim} shows that the transitional regions of $n_e$ for the
$3d^4$ levels in Ti-like ion are very close, and therefore the intensity
ratios for the observed M1 lines do not exhibit significant $n_e$-dependence.
We only present a single line ratio 16.890/13.543 (dashed line in the bottom
left panel of Fig.~\ref{Fig6}) which only changes within a factor of 3 within
the discussed range of $n_e$.

\subsection{Sc-like W$^{53+}$}

The two line ratios for the Sc-like ion, 18.867/12.312 and 17.216/12.312, show
significant variation between 10$^{14}$ cm$^{-3}$ and 10$^{16}$ cm$^{-3}$. The
former ratio is not monotonic, as is seen from Fig.~\ref{Fig6}: the
low-density limit of about 2 drops to approximately 1.3 before climbing to the
value of 10 at the high-density limit. This dip results from the presence of 
other metastable levels of $3d^3$ that interact with the upper levels of those
lines.

\section{Conclusions}

In this paper we presented measurements and identifications of forbidden
magnetic-dipole lines within ground configurations of all $3d^n$ ions of
tungsten, from Co-like W$^{47+}$ to K-like W$^{55+}$. Two sets of EUV spectra
between 10 nm and 20 nm, independently measured two years apart, excellently
agreed with each other, thereby confirming a very good reproducibility of the
results. A total of 37 new spectral lines were identified in the spectra. 

The identification of the observed M1 lines was based on extensive
collisional-radiative modeling of spectra from non-Maxwellian plasma of EBIT.
We introduced a new scheme for level grouping in order to reduce the total
number of states in our CR model to a tractable level. The calculated spectra
agree very well with the measured ones, thereby providing unambiguous
determination of line identifications.

The importance of the measured magnetic-dipole lines for spectroscopic
diagnostics of hot plasmas stems from several factors. First, the lines are
located within a rather narrow range of wavelengths, which facilitates their
measurements and reduces dependence on variation of spectrometer efficiency
with wavelength. Second, most of the lines are well isolated with only a few
overlapping to some degree. Third, the intensity ratios of spectral lines from
different ions can be used to infer electron temperature and ionization
balance over a large range of plasma density. Finally, as was shown here with
a detailed collisional-radiative modeling of Maxwellian plasmas, a large
number of line intensity ratios are sensitive to electron density in the range
of magnetic fusion devices. All these features make the M1 lines in $3d^n$
ions of tungsten especially useful for plasma diagnostics.

\begin{acknowledgments}

We thank J.M. Pomeroy, J.N. Tan, and S.M. Brewer for assistance during the
first experimental phase of this work. This work is supported in part by the
Office of Fusion Energy Sciences of the U.S. Department of Energy and by the
Research Associate Program of the National Research Council.

\end{acknowledgments}

\bibliography{w}{}

\end{document}